\documentstyle[12pt,epsfig]{article}
\begin{document} 
\vspace{15mm}
\centerline{\Large\bf Testing Quantum Mechanics in Neutrino Oscillation}
\bigskip
\centerline{Fengcai Ma$~^1$~~and~~Haiming Hu$~^{2,3}$}
\centerline{\sl $~^1$~Department of Physics, Liaoning University, 
	  Shenyang 110036, P.R.China}
\centerline{\sl $~^2$~CCAST (World Laboratory), Beijing 100080, P.R.China} 
\centerline{\sl $~^3$~Institute of High Energy Physics, Academia Sinica,}
\centerline{\sl P.O.Box 918, Beijing 100039, P.R.China}
\bigskip

\begin{abstract}
A scenario of testing quantum mechanics in neutrino oscillation is presented. The
quantum mechanics violation(QMV) that is motivated by arguments based on quantum 
gravity is investigated in neutrino system. It is found that the evolution  
equation of density matrix including QMV effect is analytically resolvable  
for neutrino propagate in vacuum or in matter under adiabatic approximation. 
The analytical formulas  have been derived.  Some bounds on the related  
parameters have been  obtained from neutrino experiments.
\end{abstract}

PACS number(s): 14.60.Pq, 03.65.Bz   
\vskip 1cm

\section{Introduction}

  It is most essential and important issue that whether neutrino has 
nonzero mass and whether there is neutrino oscillation because which is  
related to the theoretical structure of standard model (SM) of particle  
physics and is also related to astrophysics and cosmology. Recent  
experiments\cite{SN}-\cite{LSND} have shown that there are signals of  
neutrino oscillation. If it is confirmed, neutrino system may not only be  
a window beyond SM but also be a fruitful system for testing quantum  
mechanics and probing discrete symmetries of Nature as in $K^0\bar{K^0}$  
system.

The suggestion that quantum coherence might be lost at the microscopic level 
was made by Hawking\cite{Hawking}, which suggested that asymptotic  
scattering should be  described in term of a superscattering operator 
$\not S $, relating initial and final density matrices, that does not  
factorize as a product of $S-$ and $S^{+}-$ matrix elements 
\begin{equation}
  \rho_{out}=\not{S} \rho_{in},~~~~~~\not{S}\not=SS^{+} .
\end{equation}
The loss of quantum coherence was thought to be a consequence of microscopic 
quantum- gravitational fluctuations in the space- time background. Ellis,  
Hagelin, Nanopoulos and Srednicki (EHNS)\cite{EHNS} then pointed out that 
if Eq.(1) is correct for asymptotic scattering, there should be a  
corresponding effect in the quantum mechanics Liouville equation that 
describes the time evolution of the density matrix $\rho(t)$
\begin{equation}
\partial_t\rho (t)=i[\rho ,H]+i\delta \not{H}\rho
\end{equation}
The extra term may evolve a pure state into mixed state and result in 
quantum mechanics violation (QMV). The $\delta\not{H}$ was parameterized  
for simple two level system by EHNS\cite{EHNS} and was used in neutron and 
$K^0\bar{K^0}$ systems. Then the QMV was through studied by the authors  
of references\cite{Ellis92},\cite{Huet},\cite{Ellis96} for $K^0\bar{K^0}$  
system,  and by the authors of Ref.\cite{BB} for $B\bar{B}$ system. Some  
restrictions on QMV   parameters have been given. 

    A beam of two flavor neutrinos is a simple two-state system. If mass 
eigenstate is not degeneracy with weak interaction eigenstate, then mixing 
between two flavor will exist and neutrino oscillation will  occur. As in 
$K^0\bar{K^0}$ system, the quantum mechanics violation originate from   
quantum gravitational fluctuation may also exist in neutrino system. We   
investigate QMV in neutrino system in term of the formalism proposed by  
EHNS\cite{EHNS} and find that neutrino oscillation probability can be  
modified by QMV effect. Thus the neutrino oscillation experiments may present  
a precise testing for quantum mechanics. 

    The main goal of this paper is that establish a framework of QMV in  
neutrino system, derive the detailed formulas of neutrino oscillation  
probability  including QMV effect, and present some bounds on QMV parameters.  
For completeness we will first derive the formulas of  neutrino oscillation  
probabilities described by  density matrix formalism within conventional  
quantum mechanics in section II.  The QMV for neutrino oscillation in vacuum  
and in matter will be investigated  in sections III and IV respectively.   
We will apply the formulas derived in  previous section to current neutrino  
experiments and draw out certain  restrictions on QMV parameters in section V.   
Our conclusions are  presented in  section VI.

\section{Neutrino oscillation described in term of density-matrix formalism}

We now study a system consisted of two-flavor neutrinos in vacuum. The
Hamiltonian of this system, in mass eignstate, is diagonal, $H=diag(E_1,E_2)$,
here $E_1$ and $E_2$ are energy eigenvalues of neutrinos.  In some practical  
problems, for instance solar neutrino detection or atmospheric neutrino  
observation, $m_j\ll E_j$ are always valid, then  
$E_i\approx E+\frac{m_i^2}{2E},~~i=1,2$, are good approximations, here  
$E\equiv\mid p\mid$ is the magnitude of three momentum of neutrinos.   We   
describe the time evolution of the neutrino system  by Liouville equation in  
conventional quantum mechanics  
\begin{equation}
\frac{d}{dt}\rho (t)=i[\rho ,H],
\end{equation}
where $\rho$ is the density matrix of the system. Following EHNS[7], we  
expand the Hamiltonian and density matrix in Pauli matrix basis 
\begin{equation}
H=\frac{1}{2}h_{\alpha}\sigma_{\alpha},~~~ 
\rho=\frac{1}{2}\rho_{\beta}\sigma_{\beta},~~\alpha , \beta =0,1,2,3, 
\end{equation} 
where $\sigma_0$ is unit matrix and $\sigma_j~(j=1,2,3)$ are Pauli matrices. 
The evolution of the components of desity matrix obey  
\begin{equation}
\partial_t\rho_{\alpha}=h_{\alpha\beta}\rho_{\beta} . 
\end{equation}
We assume that there are only electron neutrinos in the system at the moment  
$t=0$. Considering this initial condition and  solving the differential  
equation of Eq.(5), we get the density matrix 
\begin{equation} 
\rho (t)=\left ( 
\begin{array}{lcr}
cos^2\theta & \frac{1}{2}sin2\theta e^{i\frac{\triangle}{2E}t} \\
\frac{1}{2}sin2\theta e^{-i\frac{\triangle}{2E}t}  & sin^2\theta
\end{array}\right ) , 
\end{equation}
where the $\theta$ is the vacuum mixing angle of neutrinos. The probability  
that electron neutrino oscillate to muon neutrino is given by computing the  
expection value of $Tr(O\rho )$ of an observable
\begin{equation}
O(\nu_{\mu})=\left (
\begin{array}{lcr}
sin^2\theta & -\frac{1}{2}sin2\theta \\
-\frac{1}{2}sin2\theta & cos^2\theta
\end{array}\right ) . 
\end{equation}
It is easy to check that the oscillation probability
\begin{equation}
P(\nu_e\rightarrow\nu_{\mu})=Tr\left[O(\nu_{\mu})\rho (t)\right]
\end{equation}
is consistent with that obtained from solving propagation equation of
neutrino\cite{671}.

    If neutrino propagate in matter, the Hamiltonian is corrected  
as\cite{Matter} 
\begin{equation}
\bar{H}=E+\frac{m_{2}^{2}-m_{1}^{2}}{4E}
      -\frac{1}{\sqrt{2}}G_{F}N_{n}+\frac{1}{2}\bar{M^2} , 
\end{equation}
where
\begin{equation}
\bar{M^2}\equiv\frac{1}{2}\left (
\begin{array}{lcr}
-\triangle cos2\theta +2A & \triangle sin2\theta \\
\triangle sin2\theta      & \triangle cos2\theta
\end{array}\right ),~~~
A=2\sqrt{2}G_{F}N_{e}E,
\end{equation}
the $N_e$ and $N_n$ are the number densities of electrons and neutrons in
matter. The Hamiltonian can be diagonalized by an unitary matrix 
$\bar{U}(\bar{\theta})$, here the $\bar{\theta}$ is effective mixing angle  
of neutrinos in matter,  which is determined by 
\begin{equation}
 tan2\bar{\theta}=\frac{\triangle sin2\theta}{\triangle cos2\theta -A}.
\end{equation}
The diagonalized Hamiltonian can be expanded in Pauli matrix basis as  
\begin{equation}
\bar{H_{d}}=\frac{\Sigma}{2}\sigma_{0}-\frac{1}{2}\lambda\sigma_3 ,
\end{equation}
here
\begin{equation}
\Sigma\equiv 
 2[E-\frac{1}{\sqrt{2}}G_{F}N_{n}+\frac{1}{4E}(m_1^{2}+m_2^{2}+A)],
\end{equation}
\begin{equation}
\lambda \equiv\frac{1}{2E}
\sqrt{(\triangle cos2\theta -A)^{2}+{\triangle}^{2} sin^{2}2\theta} . 
\end{equation}
In this situation, the $h_{\alpha\beta}$ in Eq.(5) becomes
\begin{equation}
h_{\alpha\beta}= \left (
\begin{array}{lccr}
0 & 0          &   0         & 0 \\
0 & 0          &  \lambda  & 0 \\
0 &  -\lambda  &   0       & 0 \\
0 & 0          &   0       & 0 
\end{array} \right ) . 
\end{equation}
In general, it is difficult to solve analytically the evolution equation of
the components of density matrix, since the $\lambda$ is a complicated
function of time. However, it is shown (see Appendex A) that if adiabatic
condition\cite{adia} is valid, then the approximation
\begin{equation}
|\frac{1}{\lambda}\frac{d\lambda}{Edt}|\ll 1
\end{equation}
is available, the term being proportional to $({d\lambda}/{dt})$  can be  
neglected during solve the differential equation of density matrix.  
The evolution equation   of density matrix can be analytically solvable.  
We still assume that there   is only electron neutrino in system and the  
effective mixing angle is  $\bar{\theta}$ at the moment $t=0$. Under this   
initial condition the desity matrix can be  evaluated out 
\begin{equation}
\rho (t)=\left (
\begin{array}{lr}
cos^{2}\bar{\theta} & \frac{1}{2}sin2\bar{\theta}exp(i\int_0^{t}\lambda dt') \\
\frac{1}{2}sin\bar{\theta}exp(-i\int_0^{t}\lambda dt') & sin^{2}\bar{\theta}
\end{array} \right ) . 
\end{equation}
If neutrinos are detected in vacuum,  the oscillation probability is 
calculated as 
\begin{equation}
P(\nu_{e}\rightarrow\nu_{\mu})=\frac{1}{2}(1-cos2\theta cos2\bar{\theta})
-\frac{1}{2}sin2\theta \sin2\bar{\theta} cos(\int_0^{t}\lambda dt') . 
\end{equation}
Which is just the well known MSW solution\cite{MSW} of solar neutrino problem. 
~~\\
\section{QMV for neutrino oscillation in vacuum}

We now introduce QMV in neutrino oscillation. Following EHNS\cite{EHNS} the  
extra term in Eq.(2) may be parameterized as a symmetrical $4\times 4$ matrix 
$h'_{\alpha\beta}$, then the evolution equation of the components of density 
matrix is rewritten as 
\begin{equation}
\partial_{t}\rho_{\alpha}=(h_{\alpha\beta}+h'_{\alpha\beta})\rho_{\beta} . 
\end{equation}
The matrix $h'_{\alpha\beta}$ must obey some restrictions. The probability 
is conserved, entropy must be real and never decrease, which imply that 
$h'_{\alpha 0}=h'_{0\beta}=0$ and $h'_{\alpha\beta}$ is negative  
semidefinite. The energy conservation in neutrino oscillation demands that 
$h'_{3 0}=h'_{0 3}=0$ have to be imposed. The $h'_{\alpha\beta} $ is 
thus parameterized as 
\begin{equation}
h'_{\alpha\beta}=\left (
\begin{array}{lccr}
0 &     0     &     0     & 0 \\
0 &  -\alpha  &  -\beta   & 0 \\
0 &  -\beta   &  -\gamma  & 0 \\
0 &     0     &     0     & 0
\end{array} \right ) , 
\end{equation}
$$
\alpha >0, ~~~ \gamma >0, ~~~ \alpha\gamma >\beta^2 . $$
By solving the Eq.(19)  we get the density matrix  
\begin{equation}
\rho (t)=\left (
\begin{array}{lr}
cos^{2}\theta &\frac{1}{2}sin2\theta (\rho_{1}-i\rho_{2}) \\
\frac{1}{2}sin2\theta (\rho_{1}+i\rho_{2}) & sin^{2}\theta
\end{array} \right ) , 
\end{equation}
where the $\rho_1$ and $\rho_2$ have different forms due to the magnitude
relation between the difference of neutrino mass square and the QMV
parameters, which will be discussed as follows.\\

(i)~~If the condition
\begin{equation}
(\alpha -\gamma)^{2} +4\beta^{2} < \frac{\triangle}{E}
\end{equation}
is satisfied, then $\rho$ has a oscillation-like solution
$$
\rho_{1}=cos\delta_{1}t~e^{-\frac{\alpha+\gamma}{2} t} , 
$$
\begin{equation}
\rho_{2}=\frac{1}{\frac{\triangle}{2E}-\beta}
   (\frac{\alpha-\gamma}{2}cos\delta_{1}t-\delta_{1}sin\delta_{1}t) 
    e^{-\frac{\alpha+\gamma}{2} t} , 
\end{equation}
where
\begin{equation}
\delta_{1}\equiv\frac{\triangle}{2E}
 \sqrt{1-\frac{(\alpha-\gamma)^{2}+4\beta^{2}}{(\triangle/E)^2}}. 
\end{equation}
The oscillation probability is 
\begin{equation}
P(\nu_{e}\rightarrow\nu_{\mu})=\frac{1}{2}sin^{2}2\theta 
(1-cos\delta_{1}t\cdot e^{-\frac{\alpha+\gamma}{2} t}) . 
\end{equation}
~~\\

(ii).~If the relation
\begin{equation}
 (\alpha-\gamma )^{2}+4\beta^{2} \ge \frac{\triangle}{E} 
\end{equation}
is valid, the $\rho$ has an exponential-like solution 
$$
\rho_{1}=[\exp{(\frac {\delta_{2}}{2} t)}+ \exp{(-\frac{\delta_{2}}{2} t)}]  
	 e^{-\frac {\alpha+\gamma}{2} t} , 
$$
\begin{equation}
\rho_{2}=\frac{1}{\frac{\triangle}{E}-2\beta} 
[(\alpha -\gamma+\delta_{2})exp{(\frac{\delta_{2}}{2} t)}+ 
(\alpha-\gamma-\delta_{2})exp{(-\frac{\delta_{2}}{2} t)}] 
 e^{-\frac{\alpha+\gamma}{2}t} , 
\end{equation}
here
\begin{equation}
\delta_{2}\equiv 
\sqrt{(\alpha-\gamma )^{2}+4\beta^{2}-(\frac{\triangle}{E})^2} . 
\end{equation}
The probability of neutrino oscillation becomes
\begin{equation} 
P(\nu_{e}\rightarrow\nu_{\mu})=\frac{1}{2}sin^{2}2\theta  
[1-\frac{1}{2}(exp{(-\frac{\alpha +\gamma -\delta_{2}}{2} t)}+ 
exp{(-\frac{\alpha+\gamma +\delta_{2}}{2}t)})] . 
\end{equation}
It is easy to show that $Tr\rho^2\ne 1$, the pure state may evolve into mixed  
state. With the $\alpha ,\beta ,\gamma\rightarrow 0$, the $\rho (t)$ return 
to  Eq.(6) in these two situations, quantum mechanics recovers.  
~~\\

\section{QMV for neutrino oscillation in matter}
~~\\

    In the situation of neutrino oscillation in matter, we add the 
parameterized $QMV$ term of Eq.(20) into the Hamiltonin of Eq.(15) and get  
the evolution equation of the components of density matrix
\begin{equation}
\partial_{t}\left (
\begin{array}{lccr}
\rho_0 \\
\rho_1 \\
\rho_2 \\
\rho_3
\end{array}\right )=
\left (
\begin{array}{lccr}
0  &     0         &      0         & 0 \\
0  & -\alpha       & \lambda -\beta & 0 \\
0 &-\lambda -\beta & -\gamma        & 0 \\
0 &      0         &      0         & 0 
\end{array} \right )
\left (
\begin{array}{c}
\rho_0 \\
\rho_1 \\
\rho_2 \\
\rho_3
\end{array}\right )
\end{equation}
Taking the adiabatic approximation and the initial condition that only  
electron neutrinos exist in the system into account , solving differential  
equations Eq.(30), we obtain the density matrix
\begin{equation}
\rho (t)=\left (
\begin{array}{lr}
 cos^{2}\bar{\theta} & \frac{1}{2}sin2\bar{\theta}(\rho_{1}-i\rho_{2}) \\
 \frac{1}{2}sin2\bar{\theta}(\rho_{1} +i\rho_{2}) & sin^{2}\bar{\theta}
\end{array}\right ) , 
\end{equation}
where $\bar{\theta}$ is the effective mixing angle of neutrinos in matter at  
the initial moment. The $\rho_1$ and $\rho_2$ are given as below.\\ 

(i).~~In the  situation of 
\begin{equation}
(\alpha -\gamma )^{2}+4\beta^{2} < 4\lambda^{2} , 
\end{equation}
then
$$
\rho_{1}=cos(\int_{0}^{t}\delta_{3} dt')\cdot e^{-\frac{\alpha +\gamma}{2} t} , 
$$
\begin{equation}
\rho_{2}=\frac{1}{\lambda -\beta}
  [\frac{\alpha -\gamma}{2} cos(\int_{0}^{t}\delta_{3}dt')-
  \delta_{3} sin(\int_{0}^{t}\delta_{3} dt')]
   e^{-\frac{\alpha +\gamma}{2} t} , 
\end{equation}
where
\begin{equation}
\delta_{3}\equiv \frac{1}{2}
   \sqrt{4\lambda^{2}-(\alpha -\gamma )^{2}-4\beta^{2}} . 
\end{equation} 
The neutrino oscillation probability is calculated as 
\begin{equation}
P(\nu_{e}\rightarrow\nu_{\mu})=\frac{1}{2}
   (1-\cos 2\theta \cos 2\bar{\theta})- 
  \frac{1}{2}sin2\theta sin2\bar{\theta}\cdot cos(\int_{0}^{t}\delta_{3} dt')
  e^{-\frac{\alpha +\gamma}{2} t} . 
\end{equation} 

(ii)~~In another situation
\begin{equation}
(\alpha -\gamma )^{2} +4\beta^{2} \ge 4\lambda^{2} , 
\end{equation}
then
$$
\rho_{1}=\frac{1}{2} cosh{(\int_{0}^{t}\delta_{4} dt')} 
  e^{-\frac{\alpha +\gamma}{2} t} , 
$$
\begin{equation}
\rho_{2}=\frac {1} {2(\lambda -\beta)} 
[(\alpha -\gamma+\delta_{4} )exp{(\int_{0}^{t}\delta_{4}dt')}+
 (\alpha -\gamma-\delta_{4} )exp{(-\int_{0}^{t}\delta_{4}dt')}]
e^{-\frac{\alpha +\gamma}{2} t} , 
\end{equation}
where
\begin{equation}
\delta_{4}\equiv \frac{1}{2} 
    \sqrt{(\alpha -\gamma )^{2}+4\beta^{2}-4\lambda^{2}} . 
\end{equation}
The neutrino oscillation probability becomes 
\begin{equation}
P(\nu_{e}\rightarrow\nu_{\mu})= 
\frac{1}{2}(1-cos2\theta cos2\bar{\theta})
 -\frac{1}{4} sin2\theta sin2\bar{\theta}
   cosh{(\int_0^{t}\delta_{4}dt')} 
     e^{-\frac{\alpha +\gamma}{2}t} . 
\end{equation}
It is not difficult to see that with the QVM parameter $\alpha , \beta , 
\gamma\rightarrow  0$, the probability $P(\nu_{e}\rightarrow \nu_{\mu})$    
back to Eq.(18) in the two situations, which is adiabatic MSW solution 
\cite{MSW} of solar neutrino problem. It is easy to calculate that  
$Tr\rho^{2}\leq 1$. In this  case, pure state may evolve into mixed state,  
and thus quantum mechanics is  violated.
~~\\
\section{An estimation for the bounds on QMV parameters}

We have derived the formulas of QMV effect in neutrino oscillation, which is 
to be tested in experiments. A complete analysis requires a detailed  
understanding  of all neutrino experiments which goes beyond the scope of this 
paper. We present here only a rough estimate of maximum order of magnitude of 
the QMV parameters as an illuminating example of testing quantum mechanics 
in neutrino oscillation. It is estimated theoretically that the maximum 
possible order of magnitude for QMV parameters $\alpha $ , $|\beta |$ , or 
$\gamma$  is $O(E^{2}/m_{Pl})$\cite{Ellis96}, which in $K^{0}\bar{K^0}$  
system is $\sim 10^{-19} GeV$, where $E$ is a typical energy scale in the  
system under discussion, and $M_{Pl}$ is the Plank energy scale. The bounds  
on these parameters from $K^{0}\bar{K^0}$ experiments have been obtained in 
Refs.\cite{Ellis92},\cite{Huet} and \cite{Ellis96}, the last one gives  
\begin{equation}
\alpha_{K^{0}\bar{K^0}} \leq 4\times 10^{-17}~GeV, ~~ 
|\beta_{K^{0}\bar{K^0}} |\leq 3\times 10^{-19}~GeV, ~~
\gamma_{K^{0}\bar{K^0}} \leq 7 \times 10^{-21}~GeV .
\end{equation}
Similary, in neutrino system, it is expected  theoretically that the order 
of magnitude of the maximum parameter is $O(E_{\nu}^{2}/m_{Pl})$, where  
$E_{\nu}$ is a typical energy scale of neutrino system. This is the order of 
$\sim 10^{-22} GeV$  for solar neutrinos or reactor neutrinos. In general, 
we may reasonably assume that the order of the maximum parameters 
in neutrino system is 
$O(\frac{E_{K^{0}\bar{K}^{0}}^{2}}{m_{Pl}}\cdot
(\frac{E_{\nu}}{E_{K^{0}\bar{k}^0}})^{2})\sim 
(\frac{E_{\nu}}{E_{K^{0}\bar{K}^{0}}})^{2} \alpha_{K^{0}\bar{K}^0}$,  
and that the magnitude relation among these parameters is retained in  
neutrino system, i.e.,  $\alpha \gg |\beta|\gg \gamma$.

    In the formulas of neutrino oscillation probability, the QMV parameters 
appear always in company with time $t$, for example $\alpha t$. There is no  
enough large time to have the $\alpha t\sim 1$ in current territorial 
experiments. Therefore,  we discuss here only solar neutrino experiments 
because of large neutrino propagation time. For the situation of neutrino 
oscillation in vacuum which is a possible solution of solar neutrino 
problem\cite{Vacuum}, From Eq.(25), we get 
\begin{equation}
\alpha +\gamma \le -\frac{2}{t}\ln|\frac{sin^{2}2\theta -2P}{sin^{2}2\theta}|, 
\end{equation}
where (and hereafter) the $P(\nu_{e}\rightarrow\nu_{\mu})$ is simply written  
as $P$. If the condition Eq.(26) is satisfied, we get
\begin{equation}
\alpha +\gamma >\frac{\triangle}{E} . 
\end{equation}
>From above arguments of  the order of magnitude of the maximum 
parameters  and the order of maximum parameter of $K^{0}\bar{K}^0$  
system in Eq.(40), we get an upper limit of another allowed region 
\begin{equation}
(\alpha +\gamma )_{max}\le 1.6\times 10^{-20}~GeV
\end{equation}
Two allowed regions of the $\alpha +\gamma$ are decided by Eq.(41)-(43).  

    We studied the upper limit of $\alpha+\gamma$ given by Eq.(41) vary 
with the $\sin^{2}2\theta$, the numerical results is shown in Fig.1. The 
lower and upper limits of the second allowed region are also shown in same 
figure. The neutrino oscillation probability $P$ is taken from the   
Kamiokande\cite{Kam} experimental data and the prediction of standard solar  
model (SSM)\cite{BP92}.  The region of $sin^{2}2\theta$ is given by recent  
fit\cite{SNP} of world  solar neutrino experiments. The $\triangle $ is taken  
as the center value of $\triangle\sim 6\times 10^{-11} GeV^{2}$, and  
$E\simeq 10 MeV$ is used. 

    In other situation, we discuss neutrino oscillation in matter, which is 
another possible solution (MSW effect)\cite{MSW} of solar neutrino problem. 
>From Eq.(35) we get 
\begin{equation}
\alpha +\gamma \le -\frac{2}{t}\ln|\frac{1-cos2\theta cos2\bar{\theta}-2P}
  {sin2\theta sin2\bar{\theta}}| . 
\end{equation}
>From Eq.(37) and Eq.(21), we get 
\begin{equation}
\alpha +\gamma >2\lambda . 
\end{equation}
Two allowed regions of $\alpha +\gamma$ are given by Eqs.(43)-(45) for  
the case of including matter effect .   
>From the Kamiokande experimental data, SSM predictions, as well as the 
fit\cite{SNP} of experimental data to the adiabatic MSW solution of solar 
neutrino problem, we obtain the numerical result of variation of the 
parameter  $\alpha +\gamma$ with the $sin^{2}2\theta $, which is shown  
in Fig.2. In numerical calculation, we used the center value of the  
$\triangle$ given in Ref.\cite{SNP},  
$\triangle\sim 1.6\times 10^{-5} eV^2$, and $E\simeq 10 MeV$. 
We also apply the average value of the $sin^2\bar{\theta}$ over the radius 
of the Sun. The electron distribution in the Sun is taken from Ref.\cite{Ne}. 

    It should be emphasized that our numerical result is only an example far 
from an exact analysis. The aim is to show how the information of bounds on 
QMV parameters is extracted from neutrino oscillation experiments. The value of the  
bounds are different if the different experimental data are used. A complete 
numerical analysis based on our analytical formulas is necessary. It is noted  
that the QMV for neutrino system in vacuum was discussed in Ref.\cite{Liu}. 
However, there is no analytical result and no any information about bounds on 
parameters in their disscussions. Moreover, what discussion in Ref.\cite{Liu} is the  
case of neutrino oscillation in vacuum, but the neutrino mixing parameters 
used in their calculation are taken form the values of MSW effect given in 
Ref.\cite{Hata}, there is no any room of small mixing angle to be allowed by   
experiments for vacuum oscillation solution of solar neutrino problem.

\section{ Conclusions}

     Neutrinos may be a useful system to precisely test quantum mechanics. 
the Liouville equation that describes the time evolution of density matrix 
of neutrino system can be modified by adding an extra term, which may be 
parameterized as a $4\times 4$ matrix. The modified evolution equation of 
density matrix can be solved analytically if neutrinos propagate in vacuum 
or if neutrinos propagate in non-uniform matter but adiabatic condition is 
available. The analytical expressions of neutrino oscillation probability 
with or without QMV effect have been derived. Based on theoretical analysis 
and neutrino experiment results we have extracted the bounds on the QMV 
parameters.  Two allowed region of $\alpha +\gamma$ have been obtained.  
It is expected that more precise restrictions on QMV parameters may be  
obtained from future neutrino experiments and complete numerical analysis. 

\vskip 1cm

{\bf  Acknowledgments} 

    We would like to thank professor Zhao-Xi Chang for useful discussions. 
The work was supported partly by the Liaoning Science  Foundation.

\vskip 1cm
\appendix
{\bf Appendix  A:~~ Generalized adiabatic condition}

    In this appendix we will demonstrate that if the adiabatic condition is 
valid the term being proportional to $ ({d\lambda}/{dt})$ can be neglected 
when we solve the evolution equation of the components of density matrix .  
>From the  Eq(14) we get 
$$
  \frac{1}{\lambda}\frac{d\lambda}{dt}=
  \frac{({A}/{\Delta})-\cos 2\theta}
  {\sqrt{(\cos 2\theta-A/\Delta)^{2}+\sin^{2}2\theta}}
  .\frac{1}{E\lambda}\frac{d\lambda}{dt}.
~~\eqno(A1) $$
We define two dimensionless parameters $C$ and $G$ as 
$$
  C\equiv \frac{1}{E\lambda}\frac{d\lambda}{dt} , ~~~~~\eqno(A2) 
$$
where $E$ is a typical enegy scale in the system under discussion, which is  
taken as neutrino energy in the system considered here.  
$$
  G\equiv |\frac{d\bar{\theta}}{dt}|/
    \frac{|\bar m_{2}^{2}-\bar m_{1}^{2}|}{2E} ,~~~\eqno(A3)
$$
where $\bar {m^{2}}_{1,2}$ are the eigenvalues of the matrix  
$\bar{M^{2}}$ defined in Eq.(10).  The adiabatic condition is expressed 
\cite{adia} as $G\ll 1$. From Eqs.(11) and (A1)-A3), we get 
$$
\frac {C}{G}=\frac{\Delta}{E^{2}}
  \frac{|\sin 2\theta\cos 2\theta|}{sin^{2}2\theta} .~~~\eqno(A4)
$$
For the problems related to current neutrino experiments, we estimate the 
maximum value of the righthanded of Eq.(A4). 
$\Delta_{max}\leq 10eV^{2} $ ,  $E_{min}\sim 1MeV$ , and   
$({|\sin 2\theta\cos 2\bar{\theta}|}/{\sin^{2}2\bar{\theta}})_{max}\leq 10^{3}$ ,  
We thus get  
$$
  \frac{C}{G}\leq 10^{-8} .~~~\eqno(A5)
$$
If adiabatic condition is valid, $ G\ll 1$ , we have $ C\ll 1$ . Therefore, 
we get the generalized adiabatic conditions in density matrix description 
of neutrino system. we can safely neglected the term containing 
$({d\lambda}/{dt})$ during we  solve the differential equation of density  
matrix as long as the adiabatic condition is satisfied. Under adiabatic  
approximation the analytical solution of the evolution equation can 
be obtained.

\vskip 1cm
\begin{figure}[h]
\centerline{\psfig{figure=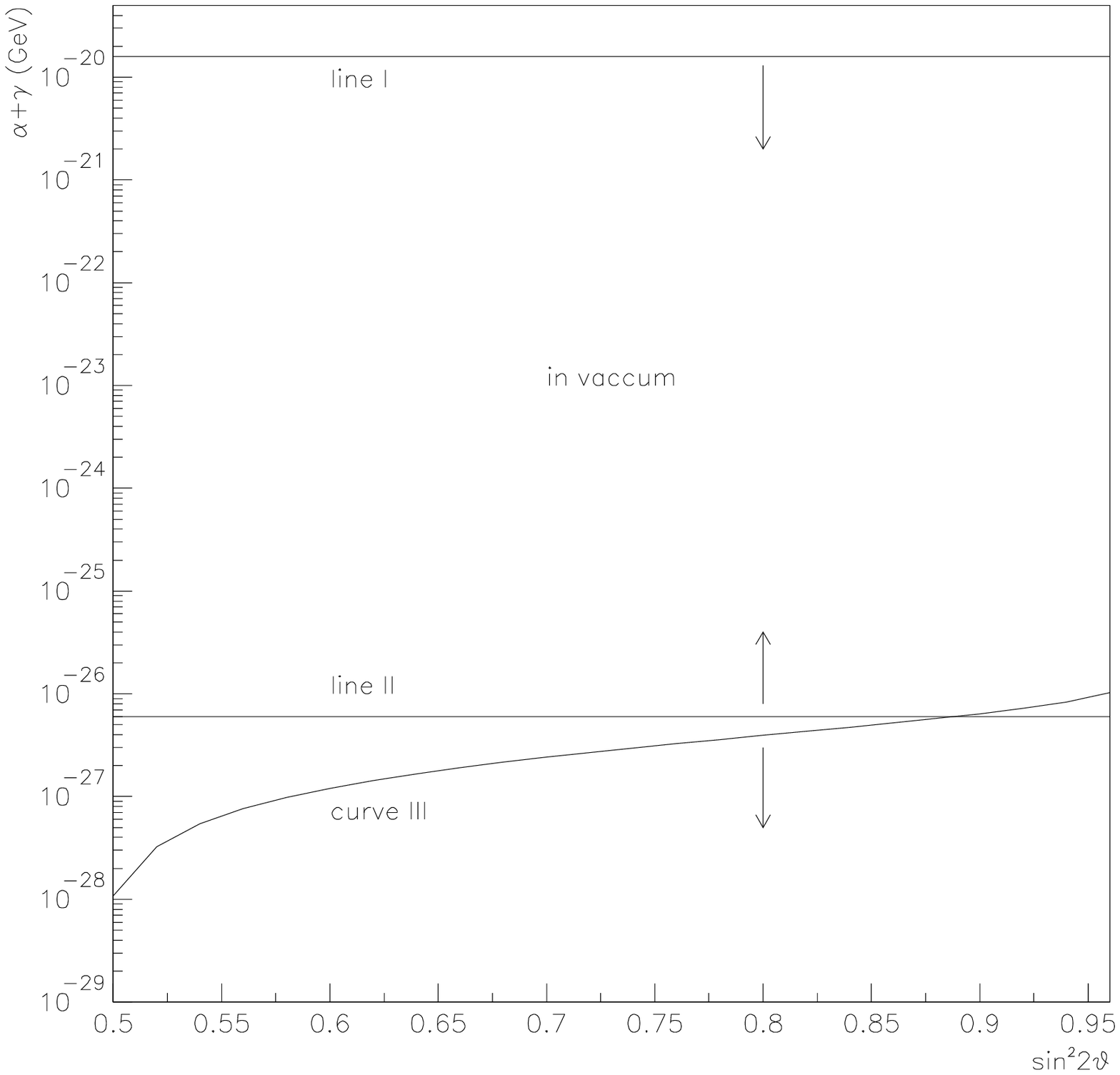,height=8cm,width=8cm}}
\caption{\label{fig1} }
\end{figure}
\begin{figure}[h]
\centerline{\psfig{figure=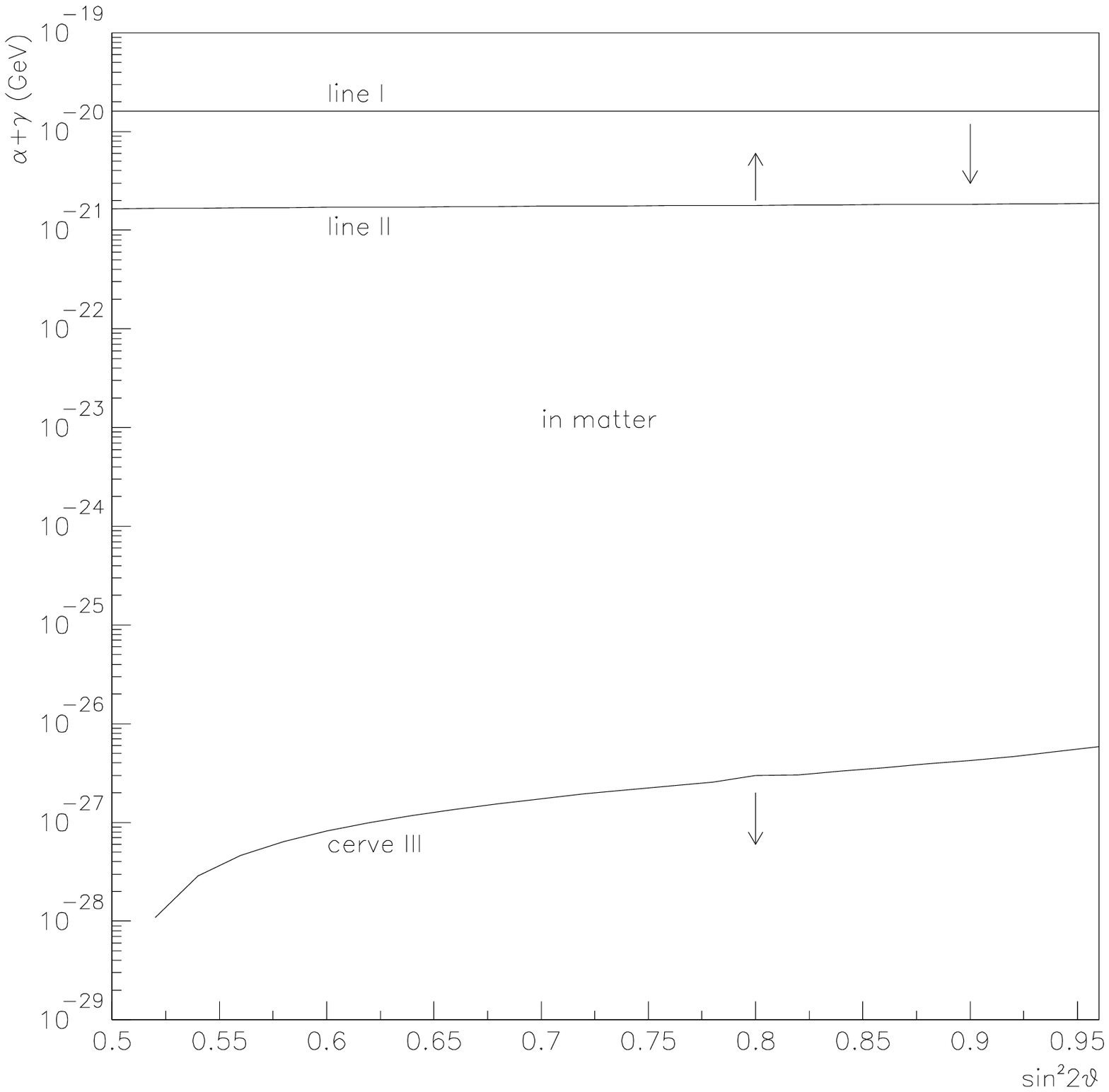,height=8cm,width=8cm}}
\caption{\label{fig2}}
\end{figure}
\newpage
\begin{center}
\large
Figure Caption
\end{center}
\normalsize
Fig.1~~The limits of the parameters $\alpha +\gamma $ vary with the 
$sin^{2}2\theta$ for neutrino oscillation in vacuum. 
The curve III is an upper  limit for the case of 
$(\alpha-\gamma)^{2}+4\beta^{2}< \frac {\Delta}{E} $. The line II is 
a lower limit for the case that above condition is not valid. 
The line I is an upper limit from the theoretical arguements and the  
 experimental upper limit of the maximum parameter in $K^{0}\bar{K}^{0}$  
 system. The arrows indicate the two allowed regions.

Fig.2~~Same as Fig.1 except for the neutrino oscillation in matter. 

\end{document}